\documentstyle{aipproc}

\begin{document}
\title{Blueshift Without Blueshift: \\
Red Hole Gamma-Ray Burst Models\\
Explain the Peak Energy Distribution 
}

\author{James S. Graber$^*$}
\address{$^*$407 Seward Square SE\\
Washington, DC 20003}

\maketitle

\begin{abstract}
Gamma-ray bursts are still a puzzle. In particular, the central engine, the total 
energy and the very narrow 
distribution of peak energies challenge model builders.  
 We consider here an extreme 
model of gamma-ray bursts based 
on highly red- and blue-shifted positron annihilation radiation.  The burst emerges from 
inside the 
red hole created by the complete gravitational collapse of the GRB progenitor.
\end{abstract}

\section*{GRB MODEL BUILDING CHALLENGES}
Because gamma-ray bursts vary so rapidly, they must be compact.  These compact gamma-ray 
bursts release 
enormous energy, and therefore they must form an intense fireball that is optically thick, 
pair-producing, 
and thermalized.  But the spectrum is not thermal, and there is no sign of 
pair-production attenuation 
at the high end of the observed spectrum\cite{jsgband}.  This seeming self-contradiction 
(the opacity problem) can be solved by having 
the fireball power a relativistic shell or jet that collides with something 
(perhaps itself) to produce 
the observed gamma rays\cite{jsgrees}.  This fireball-driven relativistic shock model is  
currently the leading candidate 
to explain GRBs\cite{jsgpiran}. It solves the opacity problem.  
But like 
almost all other published models, it fails to explain the observed spectroscopy of  
GRBs, particularly 
the narrowness of the observed peak energy distribution\cite{jsgpreece}.  Furthermore, this 
model does not 
explain the high ratio of the energy of the GRB burst itself (caused by internal shocks)  
to the energy in the afterglow (caused by external 
shocks in the fireball/shock model)\cite{jsgpa}.  
Nevertheless the predictions of this model for the afterglows themselves are consistent with 
current observations\cite{jsgpiran}. 

Finally, there is the problem of the overall energetics of the GRB.  The two leading 
candidates to 
produce the initial fireball or fireballs --the so-called central engine-- are merging 
neutron stars and 
core-collapse supernovae\cite{jsgwoosley}.  Both these sources have 
over 10$^5$$^4$ ergs of total energy 
available. 
This is more than enough energy for even the most energetic GRB, but it is not at all 
clear how to prevent most of it from 
falling into the 
newly created black hole that forms in the standard general relativity versions 
of these models.

There seems to be an inherent conflict between solving the opacity problem and solving the 
peak energy distribution problem.  The only successful technique available to solve 
the opacity problem is to invoke 
highly relativistic bulk motion.  In the relativistic frame, the gamma rays are 
below pair-production threshold and so do not 
suffer pair-production attenuation. This definitively solves the opacity problem.  
But unless the Lorentz gamma factor of 
the bulk motion can be fine-tuned to a very narrow range for all GRBs,  
the resulting blueshift will not only relocate the 
peak of the photon energy distribution; it will also substantially widen it, 
inconsistent with the observed narrow 
E-peak distribution.
Thus one needs to find a way to fine-tune the Lorentz gamma factor or find some 
other way around this conflict.  
In the fireball/shock model, the gamma factor depends sensitively on the 
baryon loading, and hence will 
vary widely.  Furthermore, the internal shocks model is dependent on shocks 
with varying Lorentz 
gamma factors  colliding with each 
other.  So narrowly limiting the gamma factor is not a reasonable option for this model.

 A generic solution to this problem is provided if the relativistic bulk motion results 
not from an initial explosion, 
but rather from the gravitational acceleration of matter falling into a deep potential well.  
An arbitrarily high 
Lorentz gamma factor can be attained, but the accompanying blueshift will be exactly cancelled 
when the matter 
and radiation are redshifted as they emerge from the potential well.  (By that time, the 
matter and radiation 
will have separated, so the opacity problem has already been solved).

A black hole can provide the necessary deep potential well. Once matter or 
radiation is deep in the 
potential well of a 
black hole, however, it is almost impossible for it to escape.  Therefore, we will consider 
an alternative 
gravitational 
collapse paradigm in which it is possible to escape from deep within the potential well 
of a gravitationally 
collapsed object. 

\section*{WHY CONSIDER RED-HOLE MODELS?}

The problems with constructing a GRB model might be sufficient motivation to consider 
alternate 
theories of gravity.  However, a stronger motivation comes from the theory of gravitation.  
Recent 
theoretical developments in string 
theory, quantum gravity and critical collapse strongly suggest the possibilities of 
both gravitational collapse 
without singularities (and without loss of information) and also gravitational 
collapse without event 
horizons\cite{jsgms,jsgst,jsgchr}.  If these 
possibilities are correct, we are forced to consider the phenomenological consequences 
(such as 
different models for GRBs and core-collapse supernovae) of alternate paradigms for 
gravitational 
collapse in which black holes do not form\cite{jsggrab}.

\section*{RED HOLES-- A NEW PARADIGM}

Many authors have considered the alternative in which a hard core collapsed object similar 
to a smaller 
harder denser neutron star forms in place of a black hole\cite{jsgrob}.   We here consider the 
alternative in 
which no such hard surface forms.  Instead the spacetime stretching that forms a black hole 
in the standard model occurs, but it does not continue to the extent necessary to form an event 
horizon or a singularity.  Instead, spacetime stretches enormously, but not infinitely, and 
forms 
a wide, deep potential well with a narrow throat.  We call this a red hole.

This type of spacetime configuration was considered by 
Harrison, Thorne, Wakano and Wheeler (HTWW)
in 1965, but only as a way station in the 
final collapse to a black hole (not yet then called by that name)\cite{jsghtww}.  
In their version, part of 
the 
configuration is inside the event horizon, the collapse continues, and a singularity soon 
forms.  

In the new alternate paradigm we call a red hole, no event horizon forms and no singularity 
forms. The gravitational collapse does not continue forever, but eventually stops.  (Why? 
Perhaps due to quantum effects or string-theory dualities, but we cannot discuss this 
adequately here.)  As the collapse proceeds, the collapsing matter becomes 
denser and denser until it reaches a critical point, after which, the distortion of spacetime 
is so great  that the density decreases.  This happens because the spacetime is stretching 
outward faster than the collapsing material can fall inward.  (This decreasing density 
effect was already noticed 
by HTWW in their analysis of gravitational collapse 
in the context of standard general relativity\cite{jsghtww}. In general relativity, 
this expansion 
of spacetime is mostly hidden behind the event horizon and does not prevent the formation of 
a singularity in a finite time. This is not the case in several observationally viable alternate 
theories of gravity\cite{jsgyil,jsgitin}.)  
This is why we are confident that the center of a red hole resembles 
a low-density vacuum more than it resembles a high-density neutron star.  The decrease in 
density due to 
this 
enormous stretching may also be a factor in halting the gravitational collapse of the red hole 
before the stretching becomes infinite.

As a result, even though the stretching of spacetime is enormous, it never becomes fast 
enough to exceed the speed of light and cause an event horizon to form.  And it stops 
before it reaches an infinite size or any other form of singularity.  (Infinite density and 
infinite curvature also do not occur.)
Nevertheless, it is very hard to escape from a red hole. First, there are trapped 
orbits inside the red hole for photons as well as massive particles, which allows permanent 
or nearly permanent trapping of mass and energy.  Second, the Shapiro delay in 
crossing a red hole is very substantial, (in some cases, enormous). Hence 
particles which are 
only crossing the red hole or passing through are in effect "temporarily" trapped.

In fact most of the matter falling into a red hole will be trapped.  However, radiation, 
and highly 
relativistic matter that falls directly into the center of the red hole and does not rescatter 
while 
inside the red hole, can travel straight through and emerge on the other side.  This 
possibility 
is essential for our proposed new GRB models. 

\section*{RED-HOLE BURST MODEL}

Elsewhere, we have considered models based on relocating part or all of the 
standard fireball/shock model inside or near a red hole. Here,  
we want to consider an even more radical model. In this model, the central engine 
is the direct 
source of the gamma-ray burst. There is no intervening finely tuned jet of baryons. 
There is no sensitive dependence on the baryon loading factor, and no dependence on 
a later shock 
to retransform the energy into gamma rays. 
Instead the 
original pair-rich fireball (created by matter collapsing into a red hole) 
becomes rapidly thin as it 
falls into the interior of the 
red hole and expands. Because everything (photons, baryons, electrons and positrons) is 
falling into the red hole at almost the same highly relativistic speed, the 
photons are below pair-production 
threshold in the infalling frame. Therefore,
the fireball is optically thin and the annhilation radiation escapes.  The 
plasma is falling with highly 
relativistic Lorentz gamma factors up to 1000 or more. The pair-annihilation photons are 
emitted in opposing pairs.  One 
is highly redshifted, while
its twin is equally and oppositely highly blueshifted.  The spectrum is highly 
broadened, but the central peak does 
not move significantly, since the net blueshift of the infalling electron-positron pair 
is balanced  by the net (or average) redshift of 
the escaping photon pair. Thus this model can solve the
narrow peak energy distribution with ease.  The more 
critical question is whether the combined annihilation line and 
thermal spectrum of the pair-rich fireball can be stretched 
enough to create the Band spectrum, or whether more 
conventional reliance on synchroton shock emission 
and/or inverse Compton scattering is necessary.

\end{document}